\documentclass[aps,prl,reprint,superscriptaddress]{revtex4-1}
\usepackage {graphicx}  
\usepackage{epstopdf}
\usepackage{bm}        
\usepackage{amssymb}   

\begin{document}

\title{Density Dependence of the Phases of the $\nu = 1$ Integer Quantum Hall Plateau \\ in Low Disorder Electron Gases}

\author{Haoyun Huang}
\author{Waseem Hussain}
\author{S.A. Myers}
\affiliation{Department of Physics and Astronomy, Purdue University, West Lafayette, IN 47907, USA}
\author{L.N. Pfeiffer}
\affiliation{Department of Electrical Engineering, Princeton University, Princeton, NJ 08544, USA}
\author{K.W. West}
\affiliation{Department of Electrical Engineering, Princeton University, Princeton, NJ 08544, USA}
\author{G.A. Cs\'athy}
\affiliation{Department of Physics and Astronomy, Purdue University, West Lafayette, IN 47907, USA}

\date{\today}
             
\begin{abstract}

Recent magnetotransport measurements in low-disorder electron systems confined to GaAs/AlGaAs samples revealed that the $\nu=1$ integer quantum Hall plateau is broken into three distinct regions. These three regions were associated with  two phases with different types of bulk localization: the Anderson insulator is due to random quasiparticle localization, and the integer quantum Hall Wigner solid is due to pinning of a stiff quasiparticle lattice. We highlight universal properties of the $\nu=1$ plateau: the structure of the stability diagram, the non-monotonic dependence of the activation energy on the filling factor, and the alignment of features of the activation energy with features of the stability regions of the different phases are found to be similar in three samples spanning a wide range of electron densities. We also discuss quantitative differences between the samples, such as the dependence of the onset temperature and the activation energy of the integer quantum Hall Wigner solid on the electron density.
Our findings provide insights into the localization behavior along the $\nu=1$ integer quantum Hall plateau in the low disorder regime.

\end{abstract}

\maketitle

The two-dimensional electron gas (2DEG) subjected to a perpendicular magnetic field supports a multitude of topological ground states. These topological ground states are characterized by topological invariants and exhibit an insulating bulk. It is widely known that bulk insulator of integer quantum Hall states (IQHSs) \cite{klitz} in low quality, high disorder samples is an Anderson-like random insulator.

The situation is very different in high quality, low disorder samples. Microwave absorption measurements of the 2DEG confined to GaAs/AlGaAs suggested that at large quasiparticle densities, these quasiparticles of the bulk may overcome disorder by ordering into a Wigner solid \cite{chen-int,lewis-1,lewis-2}. We will refer to this type of Wigner solids as the integer quantum Hall Wigner solid (IQHWS). Further evidence for the formation of the IQHWS in the GaAs/AlGaAs system is available from Knight shift \cite{muraki}, chemical potential \cite{smet}, surface acoustic wave propagation \cite{suslov}, electron tunneling \cite{ashoori}, capacitance measurements \cite{liu-cap}, and transport measurements \cite{liu-1,liu-2,sean,PRR}. In addition to the GaAs/AlGaAs system, the IQHWS was also observed in high-quality graphene \cite{young,graphene_chemical}, highlighting therefore the host-independence of this phase. To summarize, at low enough temperatures and in samples of high enough quality, experiments using various techniques found a complex behavior at the $\nu=1$ integer quantum Hall plateau. This complex behavior manifests in three regions of the plateau:
the middle region was associated with an Anderson insulator (AI), whereas in the flanks with the IQHWS \cite{sean,PRR}.

With the exception of recent results \cite{sean,PRR}, there was surprisingly little temperature-dependent transport data available in the regime in which both the AI and the IQHWS develop \cite{liu-1,liu-2}. 
The analysis of these recent transport measurements yielded a phase stability diagram \cite{sean} and established a non-monotonic, complex behavior of the activation energy \cite{PRR}. Perhaps the most puzzling topics in the physics of an integer quantum Hall plateau remain the crossover from the AI to the IQHWS and the explanation of the deep local minima of the activation energy located at the AI-IQHWS crossover. In this paper we examine these and related properties of the $\nu=1$ integer quantum Hall plateau in three samples of different densities. Our results are expected to provide quantitative inputs for theories describing localization along an integer quantum Hall plateau in 2DEGs of low disorder.

\begin{figure*}[t]
\centering
\includegraphics[width=2\columnwidth]{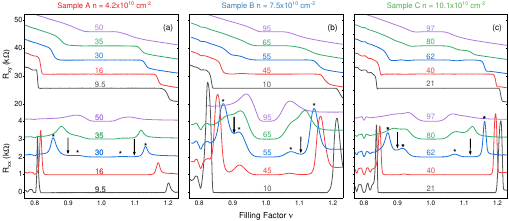}
\caption{
A representative set of longitudinal magnetoresistance $R_{xx}$ and Hall resistance $R_{xy}$ traces as plotted against the Landau level filling factor $\nu$ measured in three samples of different electron density $n$. The numerical value on top of each trace is the measurement temperature, in units of mK. Traces are offset for clarity. For the $R_{xx}$ trace marked by blue color, arrows mark reentrant integer quantum
Hall states associated with the IQHWS, while the $*$ symbols mark consecutive local maxima. These $*$ symbols establish the IQHWS-AI-IQHWS sequence of phases. Panel (b) is adapted from Ref.\cite{sean}.
} 
\end{figure*}

Samples discussed in this report belong to the newest generation of high-mobility GaAs/AlGaAs samples grown by using the most recent advances in molecular beam epitaxy \cite{chung}. These samples cover a wide electron density range, from $4.2\times10^{10}$cm$^{-2}$ to $10.1\times10^{10}$cm$^{-2}$, with sample parameters listed in Table I. Low-temperature magnetotransport measurements were performed in a dilution refrigerator using the standard lock-in technique on samples cleaved into a van der Pauw geometry. Data for the highest thermal activation energy at $\nu = 1$ of Sample B and C were measured in a Quantum Design PPMS. 

Figure 1 shows representative magnetotransport in the three samples studied in the $\nu=1$ integer quantum Hall plateau region. The horizontal axis is the Landau level filling factor $\nu$, which is calculated from
the sample density $n$ and the measured magnetic field $B$ by using the $\nu = nh/eB$ equation.
As reported in Refs.\cite{sean,PRR}, the width of the plateau, as measured in both the longitudinal resistance $R_{xx}$ and Hall resistance $R_{xy}$, is relatively narrow near $T=100$~mK and it significantly increases as the temperature is lowered. Furthermore, at certain temperatures below $T=100$~mK, such as the one associated with the blue trace in Fig.1, the plateau splinters into three distinct segments. We identify these three segments by their endpoints, i.e. the presence of local maxima in $R_{xx}$ marked by the $*$ symbols. In Refs.\cite{sean,PRR} the central region was associated with the AI and the two regions in the flanks of the plateau with the IQHWS. The latter two phases are marked in Fig.1 by vertical arrows. The transport phenomenology associated with the IQHWS is commonly referred to as the reentrant integer quantum Hall effect \cite{lilly,du, cooper,eisen02}.

\begin{table}[b]
    \centering
    \begin{tabular}{c|c|c|c}
        & Sample A & Sample B & Sample C
        \\ \hline\hline Density(cm$^{-2}$)& $4.2\times10^{10}$ & $7.5\times10^{10}$  &$10.1\times10^{10}$\\
         Mobility (cm$^2$/Vs)& $17\times10^6$& $24\times10^6$ &$35\times10^6$\\
         QW Width (nm)& $75$ & $58.5$ &$49$\\
         \hline         
         $\nu_{c,+}$ & $1.06$ & $1.06$ &$1.07$\\
         $\nu_{c,-}$ & $0.94$ & $0.93$ &$0.92$\\
         $\nu_{min,+}$ & $1.05$ & $1.06$ &$1.06$\\
         $\nu_{min,-}$ & $0.94$ & $0.93$ &$0.92$\\
    \end{tabular}
    \caption{Samples and their parameters. Here $\nu_{c,\pm}$ are the filling factors of the AI-IQHWS boundary at the lowest temperature at which the local maxima of $R_{xx}$ associated with the AI-IQHWS boundary can be observed and $\nu_{min,\pm}$ are the filling factors of the local minima of the thermal activation energy. Indexes $+$ and $-$ indicate
    electron-like and hole-like quasiparticles, respectively.
    }
    \label{sampleparameters}
\end{table}

\begin{figure*}[t]
\centering
\includegraphics[width=2\columnwidth]{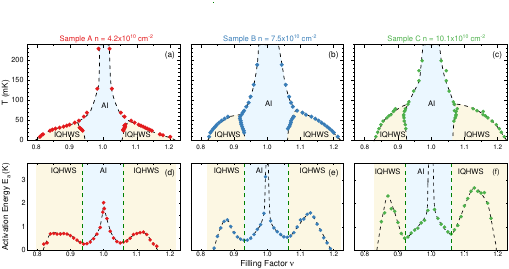}
\caption{Top panels: Phase stability diagrams in $\nu-T$ space of the three samples. The stability zone for IQHWSs is shaded yellow, whereas that for the AIs is shaded blue. The areas with a white background belong to the trivial electron liquid state. Bottom panels: Thermal activation energy $E_a$ versus $\nu$. 
Blue and yellow shaded regions correspond to the stability regions of the AI and IQHWS, as measured at the lowest temperatures at which they are detectable. Some $E_a$ data points in close vicinity to $\nu=1$ for samples B and C are off the graph. Lines are guides for the eye. Data in panel (a) are from Ref.\cite{sean} and data in panels (d) and (e) are from Ref.\cite{PRR}.
} 
\end{figure*}

The sequence of IQHWS-AI-IQHWS is observed in all three samples. To analyze the transport data, we draw the phase stability diagram in the $\nu-T$ parameter space by following the procedure outlined in Ref.\cite{sean}: at each temperature we mark the filling factors of the local maxima in $R_{xx}$. Above $T=100$~mK, there are only two local maxima; at lower temperatures there may be four.
The obtained stability diagrams are shown in Fig.2a-c. A comparison of these diagrams reveals a similar pattern across the samples: the AI, shaded in blue, is in the middle range of the plateau and it is observable across all studied temperatures. In contrast, the IQHWS, shaded in yellow, develops in the flanks of the plateau and only below a certain critical temperature. As already pointed out, perhaps the most interesting region of this stability diagram is the boundary between the AI and the IQHWS \cite{sean,PRR}. This boundary is not sharp. Indeed, the peak in $R_{xx}$ we used to identify this boundary occupies a finite range of filling factors. Therefore the AI-IQHWS boundary should be thought of as a crossover between the two phases. Furthermore, we note that at the lowest temperatures attained, the local maximum in $R_{xx}$ is no longer observable. This can be seen in the transport traces shown in Fig.1 and also in the stability diagrams shown in Fig.2.

In addition to their qualitative similarities, the stability diagrams of the three samples share quantitative aspects. The AI is not present for $\nu \gtrsim 1.1$ and $\nu \lesssim 0.9$ in any of the samples.
We find that the IQHWS of electron-like quasiparticles is stable between $\nu_{c,+} \lesssim \nu \lesssim 1.2$. Similarly, the IQHWS of hole-like quasiparticles is stable between $0.8 \lesssim \nu \lesssim \nu_{c,-}$. Here
$\nu_{c,\pm}$ represents the critical filling factor needed to stabilize the IQHWS and the indexes $+$ and $-$ indicate electron-like and hole-like quasiparticles, respectively. Values for $\nu_{c,\pm}$, as measured at the lowest temperature at which the local maximum in $R_{xx}$ is detectable at the AI-IQHWS boundary, are listed in Table.I. We notice that, aside from small variations of the order of $0.01$, the $\nu_{c,\pm}$ values of the phase boundary are density-independent. Furthermore, the stability range of the electron-like and hole-like IQHWSs is linked by the $\nu \leftrightarrow 1-\nu$ particle-hole symmetry, a pervasive property for a larger family of Wigner solids \cite{vidhi}.

The existence of a critical filling factor $\nu_{c,\pm}$ for the IQHWS can be understood as follows. As the quasiparticle density increases, quasiparticles will occupy the lowest energy bound states in the disorder potential landscape. This process has an overall smoothing effect of this potential for the subsequently added quasiparticles. As a result, the quasiparticles added past a critical density experience a lower disorder environment, increasing therefore the likelihood for the formation of charge order.

We recall that besides the Landau level filling factor $\nu$, there is another important parameter associated with quasiparticles: their areal density $n_{qp}$. In the vicinity of $\nu=1$ the two parameters are related through $n_{qp}=n |\nu-1 |$. We notice that the critical filling factor associated with the AI-IQHWS phase boundary is nearly the same in all studied samples. However, the quasiparticle density of the AI-IQHWS boundary scales with the sample density, so it differs significantly from sample to sample. 
Such a property is inconsistent with a classical electron solid and highlights therefore the quantum nature of the IQHWS. A similar finding was reported for the Wigner solid in the extreme quantum limit \cite{chen-HFWS}.

\begin{table}[b]
    \centering
    \begin{tabular}{c|c|c|c}
        & Sample A & Sample B & Sample C
        \\ \hline\hline 
        $T_{onset,+}$ (mK) & $40$ & $60$ &$80$\\
        $T_{onset,-}$ (mK) & $32$ & $60$ &$80$\\
        $E_{a,\nu=1}$ (K) & $2.0 $ & $11.6$ &$19.2$\\
        $E_{a,max+}$ (K) & $0.78$ & $1.60$ &$2.67$\\
        $E_{a,max-}$ (K) & $0.72$ & $1.30$ &$2.32$\\
    \end{tabular}
    \caption{Onset temperatures $T_{onset}$ and the maximum activation energies $E_{a,max}$ of IQHWSs of electron-like (index $+$) and hole-like (index $-$) quasiparticles.}
    \label{IQHWSstability}
\end{table}

We find that the onset temperature of the IQHWS, $T_{onset}$, i.e. the highest temperature of stability of this phase, exhibits a significant difference among the three samples. The values for $T_{onset}$ are listed in Table.II and values for electron-like quasiparticles, i.e. the highest temperature of stability for $\nu > 1.05$, are plotted in Fig.3a against the density $n$. As expected, $T_{onset}$ exhibits an increasing trend with $n$. We note that if electron-electron interactions are the sole driving force for stabilizing the IQHWS, then $T_{onset}$ is expected to scale with the Coulomb energy, thus a proportionality with $\sqrt{n}$ may be anticipated. In contrast to this expectation, data in Fig.3a exhibits a nearly linear dependence of $T_{onset}$ on $n$. This result hints at a more complex physics: disorder effects, quantum well width effects, and most likely skyrmionic effects will have to all be taken into account for a full analysis. 

The availability of temperature-dependent transport data provides access not only to the stability diagram, but also to the energy scale of low-lying thermal excitations \cite{PRR}. We found that $R_{xx}$ has a thermally activated form $R_{xx} \propto \exp{(E_a/k_B T)}$ both in the AI and the IQHWS. The extracted activation energy $E_a$ exhibits a very dramatic non-monotonic dependence on the filling factor. We found that features of the activation energy align well
with the stability regions of the different phases.
This alignment can be seen in Fig.2 for all three samples studied.

Near the mid-region of the stability range of the IQHWS, the activation energy exhibits a local maximum $E_{a,max}$. 
The values for $E_{a,max}$ are listed in Table.II and values for electron-like quasiparticles, i.e for the IQHWS at $\nu > 1.05$, are plotted in Fig.3c against the density $n$. As expected, $E_{a,max}$ exhibits an increasing trend with $n$. 
This increasing trend is expected; the skyrmionic nature of the quasiparticles of the $\nu=1$ plateau is thought to play an important role in explaining it \cite{sondhi,fertig,osw2}.  An in-depth discussion of implications of skyrmion physics on the $\nu=1$ integer quantum Hall plateau can be found in Ref.\cite{PRR}.
We note that $E_{a,\nu=1}$, measured in the center of the AI and plotted in Fig.3b, shows a similar dependence on $n$. We think that the increasing trend of both $T_{onset}$ and $E_{a,max}$ with $n$ is driven by the Coulomb energy, disorder effects, and quantum well width effects. For now, the quasi-linear relationship of $E_{a,max}$ and of $E_{a,\nu=1}$ on $n$ seen in Fig.3 remains unexplained. Quantitative disorder effects in the integer quantum Hall regime
have only been recently studied in numerical experiments \cite{osw1,osw2,osw3,osw4}.

\begin{figure}[t]
\centering
\includegraphics[width=\columnwidth]{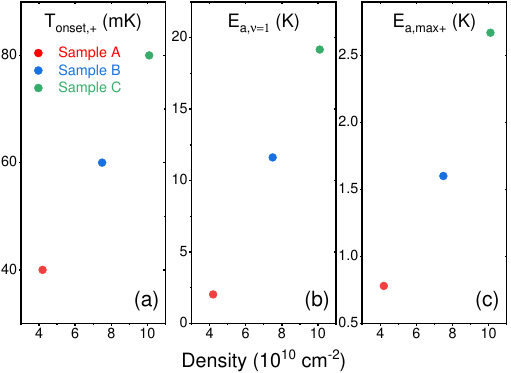}
\caption{ The dependence of (a) the onset temperature $T_{onset}$, (b) the $\nu = 1$ activation energy $E_{a,\nu=1}$ , and (c) the activation energy maxima $E_{a,max}$ of IQHWSs built of electron-like quasiparticles versus the sample density $n$. }
\end{figure}

Perhaps the most surprising result pertaining to $E_a$ is the observation of deep local minima of the $E_a$ versus $\nu$ curves at $\nu_{min, \pm}$ \cite{PRR}. As seen in Fig.2, all three samples exhibit this behavior; values of $\nu_{min, \pm}$ are listed in Table.I. We found that the filling factors of $\nu_{min, \pm}$ of the local minima in $E_a$ align  
well with $\nu_{c, \pm}$, the filling factor of the AI-IQHWS boundary. 
Furthermore, the gradual change of the activation energy near $\nu_{min, \pm}$ reinforces the existence of a crossover between the AI and the IQHWS, rather than a sharp phase transition \cite{PRR}.

We note that data from earlier microwave spectroscopy measurements in the flanks of the $\nu=1$ integer quantum Hall plateau was interpreted as evidence for two distinct IQHWSs \cite{hatke}. 
The range of filling factors associated with Wigner solid physics have a very good overlap in the two experiments, but the interpretation of the data yielded two Wigner solids in Ref.\cite{hatke} and only one in our measurements. These two results are not necessarily inconsistent;
at the lowest electron densities in Ref.\cite{hatke}, only one IQHWS was seen in the microwave data and samples in our experiments have lower densities than those accessed in Ref.\cite{hatke}.
Furthermore, the interplay of electronic ground states and disorder is not fully understood. 
For example, it is possible that an AI at high quasiparticle densities will also exhibit a signature in gigahertz microwave spectroscopy. One may also imagine scenarios in which there is only one type of IQHWS but very particular types of disorder distribution will manifest in an enhanced pinning resonance. Thus the existence of two different types of IQHWS near $\nu=1$ remains in interesting open problem. 

In conclusion, in this paper we discussed several universal properties of the $\nu=1$ integer quantum Hall plateau present in high mobility samples in a regime in which both the AI and the IQHWS develop. We found that the structure of the stability diagram, the shape of the non-monotonic dependence of the activation energy on the filling factor, and the alignment of 
features of the activation energy with features of the stability regions of the different phases
are shared properties of the three samples studied. Furthermore, we highlighted the quantum nature of the IQHWS. We also discussed trends of the onset temperature and the activation energy of the IQHWS, quantities which are driven by the density-dependent Coulomb energy.

Low temperature measurements at Purdue University were supported by the US Department of Energy award DE-SC0006671. 
The Princeton University portion of this research is funded in part by the Gordon and Betty Moore Foundation’s EPiQS Initiative, Grant GBMF9615.01 to Loren Pfeiffer.

\end{document}